\documentclass[usegraphicx,psfig,letterpaper]{mn2e}

\usepackage{amssymb}
\usepackage{times}
\usepackage{psfig}
\usepackage{amsmath}
\newread\epsffilein    
\newif\ifepsffileok    
\newif\ifepsfbbfound   
\newif\ifepsfverbose   
\newdimen\epsfxsize    
\newdimen\epsfysize    
\newdimen\epsftsize    
\newdimen\epsfrsize    
\newdimen\epsftmp      
\newdimen\pspoints     
\def\epsfbox#1{\global\def\epsfllx{72}\global\def\epsflly{72}%
   \global\def\epsfurx{540}\global\def\epsfury{720}%
   \def\lbracket{[}\def\testit{#1}\ifx\testit\lbracket
   \let\next=\epsfgetlitbb\else\let\next=\epsfnormal\fi\next{#1}}%
\def\epsfgetlitbb#1#2 #3 #4 #5]#6{\epsfgrab #2 #3 #4 #5 .\\%
   \epsfsetgraph{#6}}%
\def\epsfnormal#1{\epsfgetbb{#1}\epsfsetgraph{#1}}%
\def\epsfgetbb#1{%
\openin\epsffilein=#1
\ifeof\epsffilein\errmessage{I couldn't open #1, will ignore it}\else
   {\epsffileoktrue \chardef\other=12
    \def\do##1{\catcode`##1=\other}\dospecials \catcode`\ =10
    \loop
       \read\epsffilein to \epsffileline
       \ifeof\epsffilein\epsffileokfalse\else
          \expandafter\epsfaux\epsffileline:. \\%
       \fi
   \ifepsffileok\repeat
   \ifepsfbbfound\else
    \ifepsfverbose\message{No bounding box comment in #1; using defaults}\fi\fi
   }\closein\epsffilein\fi}%
\def\epsfsetgraph#1{%
   \epsfrsize=\epsfury\pspoints
   \advance\epsfrsize by-\epsflly\pspoints
   \epsftsize=\epsfurx\pspoints
   \advance\epsftsize by-\epsfllx\pspoints
   \epsfxsize\epsfsize\epsftsize\epsfrsize
   \ifnum\epsfxsize=0 \ifnum\epsfysize=0
      \epsfxsize=\epsftsize \epsfysize=\epsfrsize
     \else\epsftmp=\epsftsize \divide\epsftmp\epsfrsize
       \epsfxsize=\epsfysize \multiply\epsfxsize\epsftmp
       \multiply\epsftmp\epsfrsize \advance\epsftsize-\epsftmp
       \epsftmp=\epsfysize
       \loop \advance\epsftsize\epsftsize \divide\epsftmp 2
       \ifnum\epsftmp>0
          \ifnum\epsftsize<\epsfrsize\else
             \advance\epsftsize-\epsfrsize \advance\epsfxsize\epsftmp \fi
       \repeat
     \fi
   \else\epsftmp=\epsfrsize \divide\epsftmp\epsftsize
     \epsfysize=\epsfxsize \multiply\epsfysize\epsftmp   
     \multiply\epsftmp\epsftsize \advance\epsfrsize-\epsftmp
     \epsftmp=\epsfxsize
     \loop \advance\epsfrsize\epsfrsize \divide\epsftmp 2
     \ifnum\epsftmp>0
        \ifnum\epsfrsize<\epsftsize\else
           \advance\epsfrsize-\epsftsize \advance\epsfysize\epsftmp \fi
     \repeat     
   \fi
   \ifepsfverbose\message{#1: width=\the\epsfxsize, height=\the\epsfysize}\fi
   \epsftmp=10\epsfxsize \divide\epsftmp\pspoints
   \newcount\figskipcount
      \message{#1 \the\epsfysize  }
   \vbox to\epsfysize{\vfil\hbox to\epsfxsize{%
      \includegraphics{#1}%
      \hfil}}%
\epsfxsize=0pt\epsfysize=0pt}%

{\catcode`\%=12 \global\let\epsfpercent=
\long\def\epsfaux#1#2:#3\\{\ifx#1\epsfpercent
   \def\testit{#2}\ifx\testit\epsfbblit
      \epsfgrab #3 . . . \\%
      \epsffileokfalse
      \global\epsfbbfoundtrue
   \fi\else\ifx#1\par\else\epsffileokfalse\fi\fi}%
\def\epsfgrab #1 #2 #3 #4 #5\\{%
   \global\def\epsfllx{#1}\ifx\epsfllx\empty
      \epsfgrab #2 #3 #4 #5 .\\\else
   \global\def\epsflly{#2}%
   \global\def\epsfurx{#3}\global\def\epsfury{#4}\fi}%
\def\epsfsize#1#2{\epsfxsize}
\def\figinsert#1#2{\epsfbox{#1} \message{#2} }    

\begin{document}
\title[Correlations between bright submillimetre sources and
low-redshift galaxies] {Correlations between bright submillimetre sources and
low-redshift galaxies}
\author[O. Almaini et al.]
{
O.~Almaini$^{1}$, J.S. Dunlop$^{2}$, C.J. Willott $^{3}$,
D.M. Alexander$^4$, F.E. Bauer${^4}$, C.T. Liu $^{5}$ 
\\
$^1$ School of Physics \& Astronomy, University of Nottingham, University Park, Nottingham NG7 2RD \\
$^2$ Institute for Astronomy, 
University of Edinburgh, Royal Observatory, Blackford
Hill, Edinburgh EH9 3HJ \\
$^3$ Herzberg Institute of Astronomy, 5071 West Saanich Rd, Victoria, 
B.C., Canada \\
$^{4}$ Institute of Astronomy, Madingley Road, Cambridge, CB3 0HA\\
$^{5}$ 
Astrophysical Observatory, City University of New York/CSI, 2800 Victory
Blvd.,  Staten Island, NY 10314 USA}
%

\date{MNRAS in press}
\maketitle

\begin{abstract}
We present evidence for a positive angular correlation between bright
submillimetre sources and low-redshift galaxies. The study was conducted
using 39 sources selected from 3 contiguous, flux-limited SCUBA
surveys, cross-correlated with optical field galaxies with magnitudes
$R<23$ (with a median redshift of $z\simeq 0.5$).  We find that the
angular distribution of submm sources is skewed towards overdensities
in the galaxy population, consistent with $25\pm 12$ per cent being
associated with dense, low-redshift structure.  The signal appears to
be dominated by the brightest sources with a flux density
$S_{850\mu{\rm m}}>10$ mJy.  We conduct Monte-Carlo simulations of
clustered submm populations, and find that the probability of
obtaining these correlations by chance is less than 0.4 per cent.  The
results may suggest that a larger than expected fraction of submm
sources lie at $z\simeq 0.5$. Alternatively, we argue that this signal
is most likely caused by gravitational lensing bias, which may be
entirely expected given the steep submm source counts.  Implications
for future submm surveys are discussed.

\end{abstract}


\section{Introduction}

The advent of the SCUBA array at the James Clerk Maxwell Telescope led
to the discovery of a high surface density of apparently
high-redshift, dust-enshrouded galaxies (e.g., Smail et al. 1997;
Hughes et al.  1998; Barger et al. 1998, Eales et al. 1999). Progress
beyond these original discoveries has been limited however, not least
because of the optical faintness of many of these sources. Combined
with the relatively large $10-15$ arcsec SCUBA beam, this makes
unambiguous identification extremely difficult (e.g., Smail et
al. 2002).  This deadlock has recently been
partially overcome with deep radio observations at the VLA, which have
yielded precise locations for significant samples of submm galaxies
for the first time. Follow-up spectroscopy has produced the first
reliable $N(z)$ estimate for the population of SCUBA galaxies (Chapman
et al. 2003), showing a median redshift of $z=2.4$ with only a small
fraction (a few per cent) of galaxies at $z<1$.

There have been suggestions that many submm sources could be powered
by AGN (e.g., Almaini, Lawrence \& Boyle 1999), but the failure to
detect significant numbers as luminous X-ray sources suggests that the
AGN-dominated fraction is likely to be small (Fabian et al.\ 2000;
Severgnini et al. 2000; Almaini et al. 2003; Waskett et
al. 2003). Nevertheless, a large fraction (probably $> 40$\%) appear
to host moderate-luminosity AGN activity (Alexander et al. 2003). The
suggestion that some fraction could be very local cold clouds within
the Milky Way (Lawrence 2001) has not been entirely ruled out, but it
now seems likely that the majority are high redshift, dust-enshrouded
galaxies with the observed submm emission dominated by the
re-radiation of absorbed stellar light by dust.

As the true nature of these sources emerges, the next step is to make
use of this population to constrain both the history of obscured
star-formation and theories for the formation of massive galaxies.  A
crucial step will be to measure the clustering of the submm galaxies,
which will directly test the hypothesis that these are the progenitors
of massive elliptical galaxies (Percival et al. 2003). There are
preliminary indications that the SCUBA sources are clustered (Scott et
al. 2002), but surveys containing hundreds of submm sources will be
required to determine the strength of clustering to the required
precision. There is also intriguing evidence that submm sources are
clustered with Lyman-Break Galaxies (Webb et al. 2003) and X-ray
emitting AGN (Almaini et al. 2003). The first result can naturally be
explained in terms of large-scale structure at $z\sim 3$, but the
correlation with X-ray sources is more puzzling. Recent determinations
of the redshift distribution of Chandra-selected X-ray sources
(Hornschemeier et al. 2001; Gilli et al. 2003) show the majority seem
to lie at $z<1$.  This may indicate that many submm sources lie at
lower redshift than previously assumed.  An alternative possibility
(motivated by the very steep submm number counts, e.g. Scott et
al. 2002) is that these correlations are an artifact of gravitational
lensing. It is now emerging that hard X-ray sources are excellent
tracers of peaks in the density field at $z<1$ (Gilli et al. 2003;
Elbaz \& Cesarsky 2003). Could these structures lead to a substantial
gravitational lensing bias for distant submm populations?  There are
already indications that a few SCUBA sources are falsely associated
with low-redshift galaxies because of strong gravitational lensing
(Chapman et al. 2002, Dunlop et al. 2004). Combined with the more
subtle effects of weak gravitational lensing, models predict that
approximately $30$ per cent of the sources selected at $S_{850\mu{\rm
m}}>10$mJy could be gravitationally lensed (Perrotta et al. 2003; see
also Blain et al. 1999), although the precise fraction depends
critically on the slope of the (intrinsic) submm source counts.

To explore these issues further, in this paper we examine the
possibility that a significant fraction of the submm population are
correlated (in 2-D) with low-redshift large-scale structure, as traced
by optically-selected galaxies. Section 2 outlines the optical and
submm data used for this analysis. Section 3 compares the
distribution of these populations for individual and combined
fields. Section 4 examines the effects of an intrinsically clustered
submm population using Monte-Carlo simulations.  In Section 5 we
discuss the predictions of gravitational lensing bias, while Section 6
discusses the effects of cosmic variance between our fields. Section 7
presents the summary and conclusions.

\begin{figure*}
\centering
\centerline{\epsfxsize=17truecm 
\figinsert{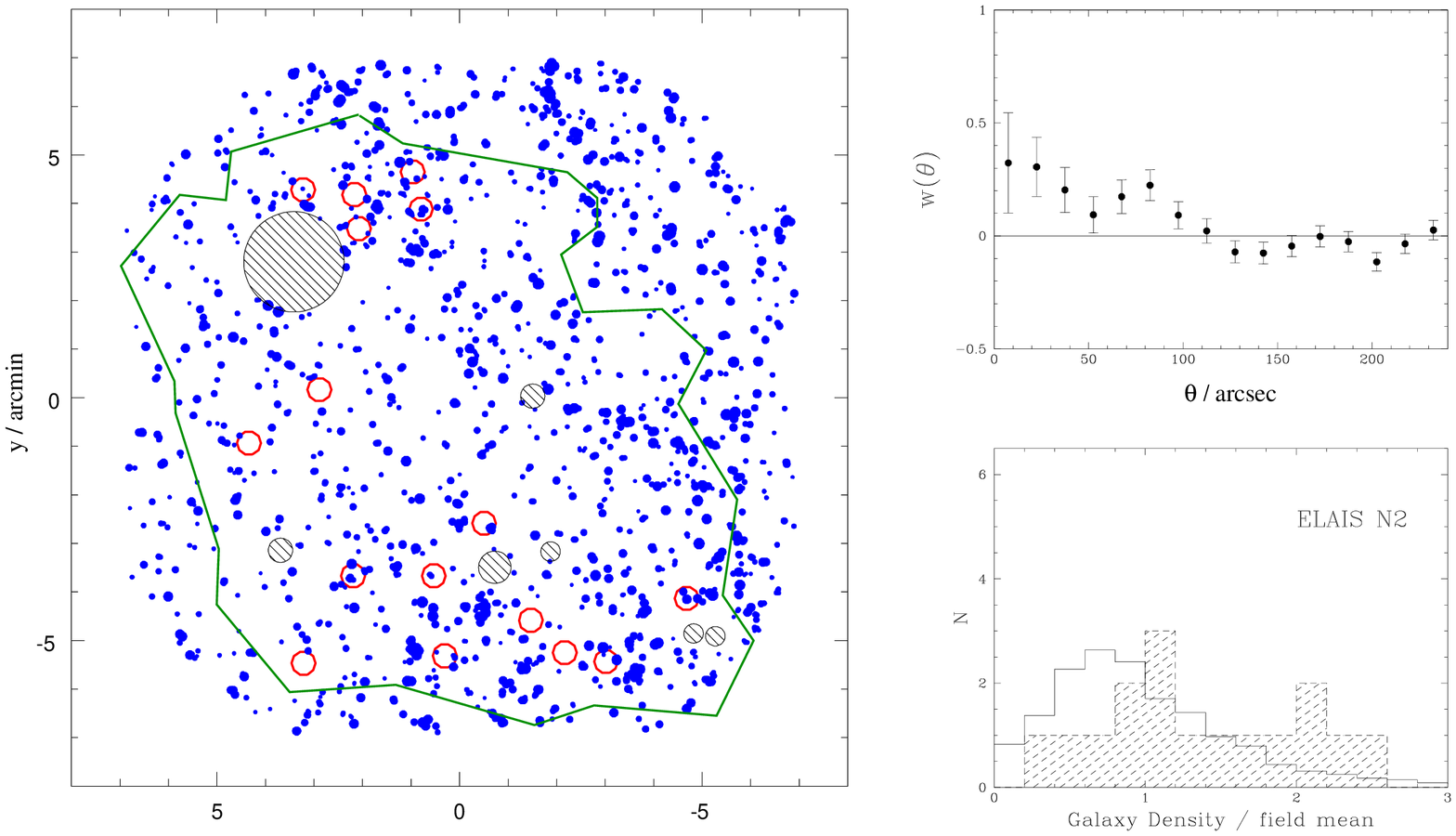}{0.0pt}}
\centering
\centerline{\epsfxsize=17truecm 
\figinsert{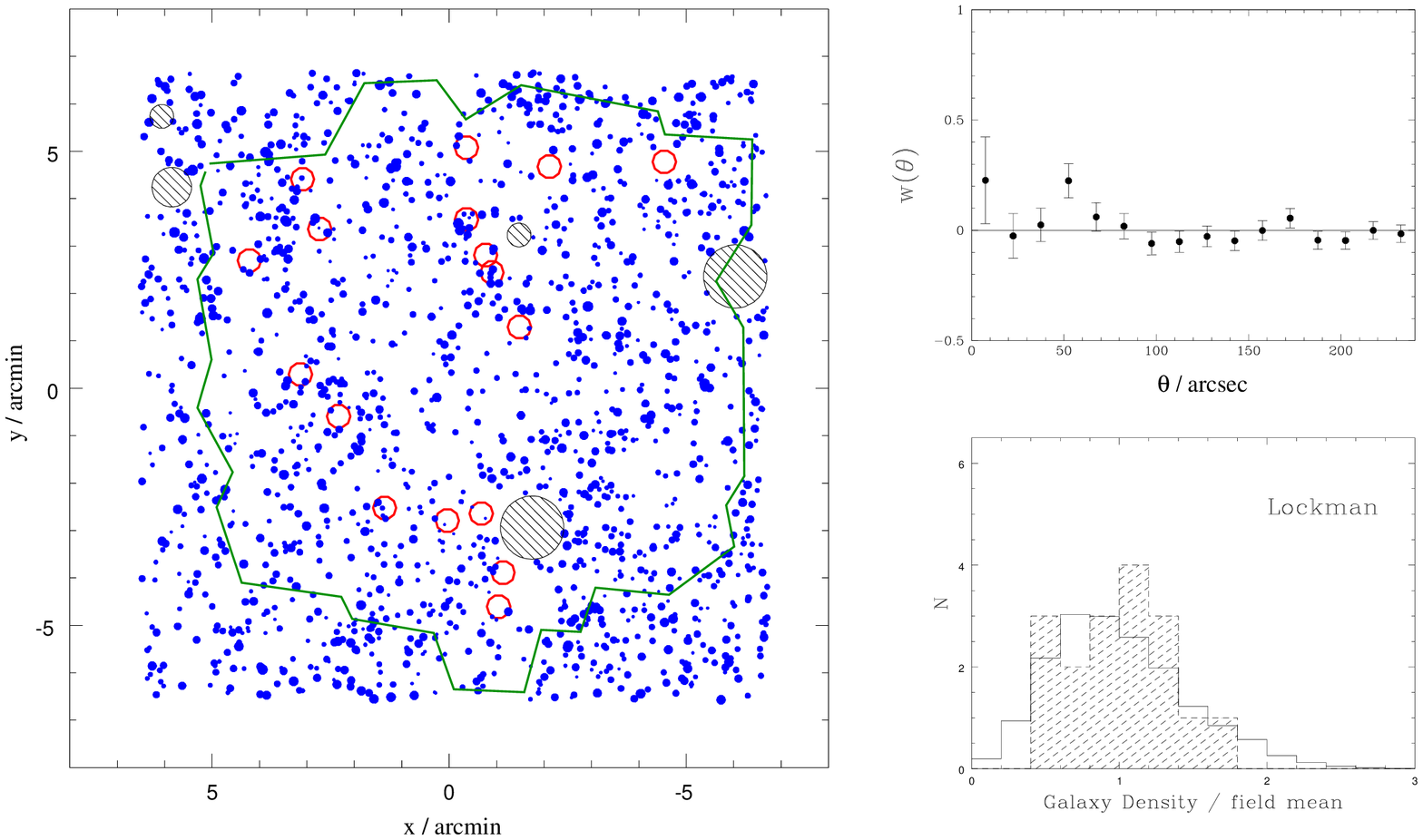}{0.0pt}}
\caption{LEFT COLUMN: The distribution of SCUBA submm sources (open
circles) and galaxies (filled points) to a limit of $R<23$ in the
ELAIS-N2 field (upper panel) and $I<22.5$ in the Lockman Hole (lower
panel). The size of the galaxy points reflects their optical
brightness. Hatched areas denote regions removed from the optical
catalogue due to the presence of bright stars.  The boundaries of the
SCUBA maps are shown by the solid line. RIGHT COLUMN: This shows the
SCUBA/galaxy cross-correlation function $w(\theta)$ and a histogram of
the galaxy densities within a 30 arcsec radius of each SCUBA source
(dashed histogram) compared to the expected distribution from 100,000
randomly placed apertures (solid histogram).}
\end{figure*}

\begin{figure*}
\centering
\centerline{\epsfxsize=17truecm 
\figinsert{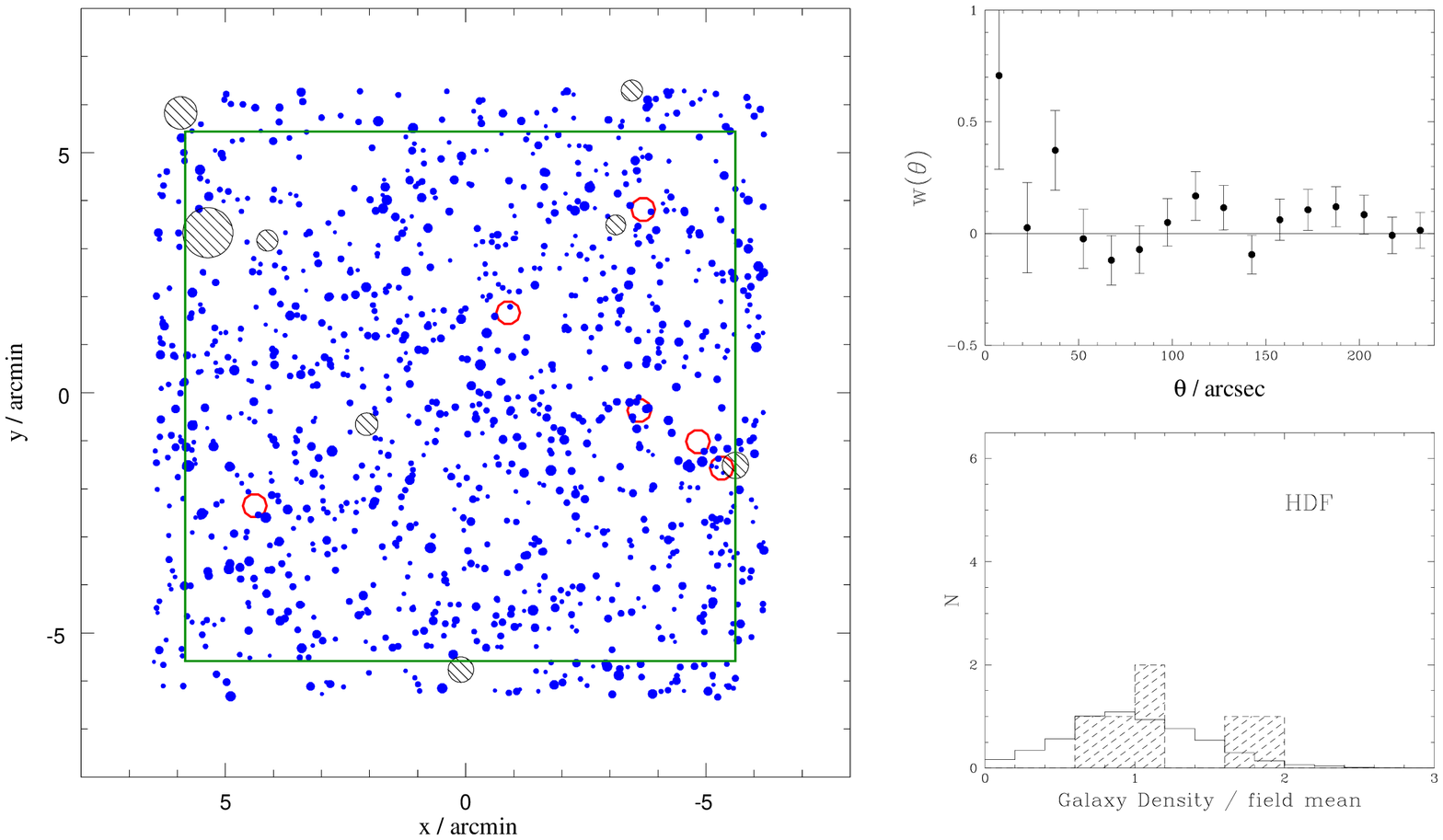}{0.0pt}}
\caption{As Figure 1, but using the submm sources from the scan-map
of the HDF region (Borys et al. 2002) relative to galaxies selected to
a limit of $R<23$.}
\end{figure*}

\section{The Observational Data}
 
\subsection{The submm surveys}

To examine any associations between submm sources and low-redshift
galaxies in an unbiased manner we must select our submm catalogue
from complete, flux-limited surveys. To this end we select 39 submm
sources detected above a S/N limit of $3.5\sigma$ from three separate
contiguous surveys. Two are from the 8mJy SCUBA survey of Scott et
al. (2002), which covers 260 arcmin$^2$ over two fields (ELAIS-N2 and
the Lockman-Hole East) using a jiggle-mapping technique to an rms
noise level of $\sim 2.5$mJy per beam at $850\mu$m. The third is from
the shallower SCUBA map of the Hubble Deep Field (HDF) conducted by
Borys et al. (2002), which covered 125 arcmin$^2$ in scan-map mode to
an rms noise level of approximately 3.5mJy.

To minimise spurious sources, we follow Ivison et al. (2002) and
exclude the sources from the 8mJy survey which lie in regions where
the noise is in excess of 3mJy per beam. This removes 5 sources from
the Lockman-Hole and 1 source from ELAIS-N2.  We also remove any
sources from the sample of Borys et al. (2002) which were not
subsequently confirmed (above $3.5 \sigma$) in the 'super-map' paper
of Borys et al. (2003).  We also exclude source
SMMJ123608+621246. This source was not confirmed in deeper
observations by Wang et al. (2004) or in a re-analysis of the same
data by Borys (private communication).

The final sample therefore contains 16 sources from ELAIS-N2, 17
sources from the Lockman Hole and 6 sources from the HDF.

\subsection{The optical galaxy catalogue}

Our aim is to determine whether the SCUBA sources are associated with 
low-redshift large-scale structure, particularly structure which could cause
gravitational lensing for an intrinsically high-redshift population.

For a background source at $z>2$, the optimal redshift for
gravitational lensing occurs for a foreground lens at $z\simeq 0.5$
(Clowe et al. 2000), assuming the currently favoured Lambda-dominated
cosmology. In the absence of redshifts for the majority of field
galaxies in the surveys regions described above we choose to
cross-correlate the SCUBA catalogues with galaxies selected from
optical photometry.

To estimate the redshift distribution of our optical galaxies we use
the photometric redshift surveys of the COMBO-17 consortium (Wolf et
al. 2003). Based on these data, we select galaxies with Vega
magnitudes $R<23$. Galaxies selected in this manner show a median
redshift of $z=0.53$ (Wolf, private communication) with only a small
tail of a few per cent at $z>1$, and as such are ideal probes of
low-redshift large-scale structure.

An $R$-band image of ELAIS-N2 was obtained with the Prime Focus Camera
(PFC) on the William Herschel Telescope (WHT) during May 1999,
primarily for the purpose of identifying X-ray sources from the ELAIS
Deep X-ray Survey (Manners et al. 2003; Willott et al. 2003). This
uses EEV detectors with a pixel scale of 0.236$''$, covering a field
of view of approximately $16\times16$ arcmin. In measured seeing of
$0.7''$ seeing, a $5\sigma$ limiting magnitude of $R=25.0$ in a $4''$
aperture is obtained. Galaxy catalogues to a limit $R<23.0$
are then extracted using the Sextractor package. Candidate stellar
sources with a CLASS\_STAR parameter $>0.8$ are removed from the
analysis.

We have no access to deep R-band imaging for the Lockman region, so we
use an I-band image also obtained using the WHT PFC during Nov 2000.
In measured seeing of $0.8''$ this reaches a $5\sigma$ limiting
magnitude of $I=23.9$.  Based on the I-band catalogues from the
COMBO-17 survey, we find that a limiting magnitude of $I=22.5$
provides a catalogue of galaxies with a near-identical $N(z)$
distribution to a sample selected with $R<23$, removing candidate
stars as above.

In the HDF region we use the catalogue of Capak et al. (2004),
obtained using the Suprimecam camera at the Subaru telescope in
February and March 2001. This reaches a $5\sigma$ limiting magnitude
of approximately $R_{AB}=26.6$. A galaxy catalogue to a depth of
$R_{\rm Vega}<23$ was then extracted using star/galaxy separation flags
provided by Capak (private communication).

Finally, the regions surrounding the brightest stars (typically those
with $R<16$) are excluded from the analysis described below to avoid
incompleteness in these regions.

\section{Correlations between optical and submm galaxies}

\subsection{Individual fields}

Figures 1 and 2 show the distribution of SCUBA submm galaxies
compared to the low-z galaxy catalogues described above.  By eye, the
ELAIS-N2 field shows a strong correlation between these
populations\footnote {Interestingly this same general structure is
seen in the Chandra X-ray population in this field (Almaini et
al. 2003), strengthening recent suggestions that hard X-ray sources
are strong tracers of peaks in the large-scale structure at $z<1$
(Gilli et al. 2003, Elbaz \& Cesarsky 2003).}.  In the Lockman and HDF
fields any correlations are visually less striking, but again there is
evidence for a number of SCUBA sources lying in the vicinity of
overdense structure.

To quantify this visual impression we perform a two-point angular
cross-correlation, using the same basic estimator defined in Almaini
et al. (2003). As expected, this reveals a clear excess of low-z
galaxies around SCUBA sources in the ELAIS-N2 field ($3-4 \sigma$
significance within 1 arcmin) with weaker evidence for an excess in
the Lockman and HDF fields ($\sim 2 \sigma$ significance within 1
arcmin).  A problem with the cross-correlation estimator, however, is
that it effectively averages out the correlations over all submm
sources. In order to estimate the fraction of submm sources lying in
overdense regions we henceforth consider the {\em distribution} of
galaxy environments traced by the submm sources. We select a $30$
arcsec radius ({\em a-priori}) around each SCUBA source and measure
the density of galaxies observed compared to the field mean.  The
resulting distributions are shown as histograms in Figures 1 \& 2. We
compare with the distribution expected from $100,000$ randomly placed
apertures distributed within the SCUBA map regions.

For the ELAIS-N2 field in particular, the submm sources appear to be
skewed towards regions of high galaxy density. A Kolmogorov-Smirnov
test (on the unbinned distributions) gives a probability of $>99$ per
cent that the SCUBA sources are not drawn from the same parent
population as the randomly placed apertures.  For both the Lockman
Hole and HDF regions the submm sources also appear slightly skewed
towards high density regions, although at much lower statistical
significance (formally rejecting the null-hypothesis at only the $85$
per cent level in each case).

\subsection{The combined data}

In Figure 3 we show a histogram of galaxy densities from the three
fields combined, formed by simply adding the distributions from the
individual fields.  We note that the peak in the expected density
distribution occurs at $\rho/\bar{\rho}<1$, reflecting the fact that
the optical galaxies are clustered. A typical $30$-arcsec aperture
will therefore sample slightly below-average density compared to the
overall field mean.

The combined data show clear evidence that the SCUBA sources are
skewed towards higher density regions in the galaxy
distribution. Formally a K-S test rejects the hypothesis that these are
drawn from the same underlying population as the random apertures at
$>99$ per cent significance.

The maximum deviation in the K-S test occurs at a density
$\rho/\bar{\rho}=0.85$ (approximately mid-way between the peaks in the two
distributions shown in Figure 3).  To estimate the {\em fraction} of
``excess'' SCUBA sources producing this signal, we consider the number
observed in regions above this density value. From a total of $39$
SCUBA galaxies we observe $30$ with $\rho/\bar{\rho}>0.85$ compared to
$20.3$ expected. This corresponds to an estimated ``excess'' of $9.7$
SCUBA galaxies, or ~$24.9\pm 11.6$ per cent of the total (assuming
Poisson statistics).

A more conservative estimate can be obtained by considering the
fraction lying in the densest regions with $\rho/\bar{\rho}>1$. In
these regions we observe $24$ SCUBA galaxies compared with $16.3$
expected, corresponding to an excess of $19.7 \pm 10.3$ per cent. The
statistical significance of these results are investigated further in
Section 4.

\subsection{Bright and faint SCUBA galaxies}

A prediction of gravitational lensing models (e.g. Perotta et
al. 2003) is that brighter submm sources are more likely to be
gravitationally lensed. This will be expected if the slope of the
(unlensed) source counts steepen, which may reflect a turn over in the
underlying submm luminosity function. Adopting the galaxy formation
models of Granato et al.  (2001), Perotta et al. (2003) predict a sharp
increase in the lensed fraction for  $S_{850\mu{\rm m}}>10$mJy.

To test this prediction we split the sample into the 13 brightest
SCUBA galaxies with $S_{850\mu{\rm m}}>10$mJy (Figure 4) and 26
fainter systems below this flux density (Figure 5).  The brighter
galaxies show a notably more skewed distribution.  Formally a K-S test
rejects the hypothesis that these are drawn from the same underlying
population at the random apertures at $>99$ per cent significance.
In contrast, a KS-test on the fainter galaxies only rejects the
null-hypothesis at $>85$ per cent significance.

We note that while the ELAIS-N2 field gave the strongest correlations
in Section 3.1, only three of these sources are brighter than 10mJy
and these do not appear to dominate the signal here (see shaded histogram
in Figure 4). Excluding these sources, the significance of the K-S
test drops, but the null hypothesis is still rejected at $>95$ per
cent significance.

We conclude that there is strong evidence for a correlation between
low-z structure and bright SCUBA galaxies.  Determining the precise
relationship as a function of submm flux density, however, will
clearly require a larger sample.

For the SCUBA galaxies with $S_{850\mu{\rm m}}>10$mJy we can estimate
the fractional ``excess'' which are associated with the low-z
galaxies. In high-density regions ($\rho/\bar{\rho} > 1$) we find 10
SCUBA galaxies, compared to only 5.4 expected from a random
distribution. Although these are small-number statistics, from the
total of 13 submm sources this corresponds to an estimated excess of
$35\pm 18$.

\begin{figure}
\psfig{figure=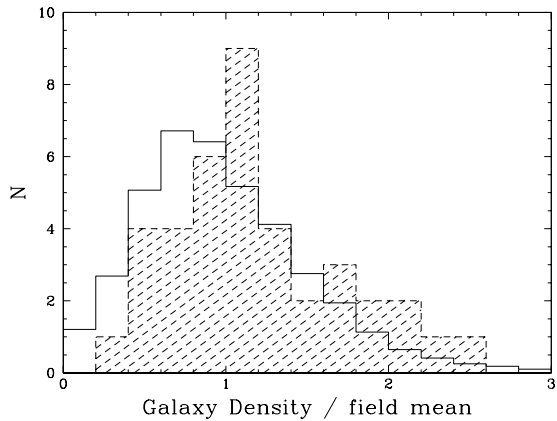,width=0.5\textwidth,angle=270}
\caption{Histograms of the optical galaxy densities within a 30
arcsec radius of the 39 SCUBA sources from three fields combined
(dashed region) compared to the expectation from randomly placed
apertures (solid line).  }
\end{figure}

\begin{figure}
\psfig{figure=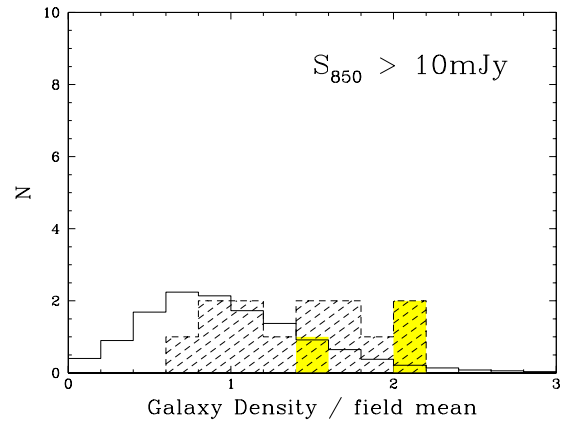,width=0.5\textwidth,angle=270}
\caption{As Figure 3, but using only the 13 brightest SCUBA
sources detected above a flux limit of $S_{850\mu{\rm m}}>10$mJy.  The
shaded sub-histogram highlights the 3 sources from the
ELAIS-N2 field.   }
\end{figure}

\begin{figure}
\psfig{figure=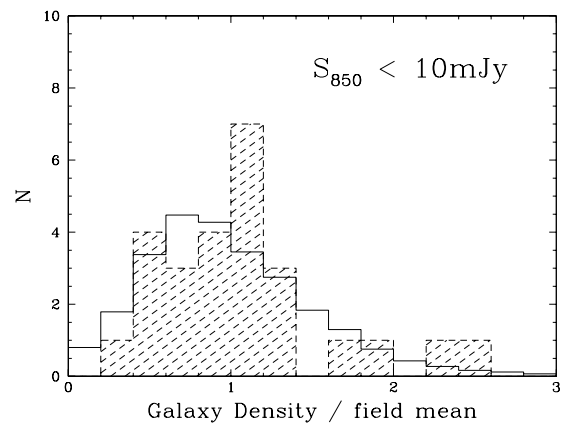,width=0.5\textwidth,angle=270}
\caption{As Figure 3, but using only the 26 fainter SCUBA sources with
$S_{850\mu{\rm m}}<10$mJy.  }
\end{figure}

\subsection{The choice of aperture size}

Throughout the density analysis we have used a $30$-arcsec radius
aperture. This corresponds to a comoving radius of $270$kpc at
$z=0.5$, which is similar to the size of a dark matter halo for a
small group or a large galaxy (Fischer et al. 2000).  For completeness
we repeated the analyses with a range of aperture sizes from $20-60$
arcsec, but found that this had no major influence on any of the
results presented here. Formally the $20$ arcsec aperture gave
marginally more significant correlations and a $60$ arcsec aperture
gave the least.  The differences were small, however, so we will not
comment further.  Hereafter we therefore stick to our {\em a-priori} choice
of $30$-arcsec radius apertures.

\section{Simulations of clustered submm sources}

An implicit assumption in the density analysis presented above is that
the SCUBA sources are intrinsically randomly distributed.  Although a
definitive measurement of submm source clustering has yet to be made
(Scott et al. 2002) in this section we explore the implications of a
submm population which is intrinsically highly clustered.  If the
SCUBA sources are strongly clustered one would expect a higher
probability of finding a large number clustered around low-redshift
overdensities (or voids) purely by chance.

Appendix A describes a technique for producing mock 2-D catalogues of
clustered populations.  We generated three sets of catalogues with
differing clustering strengths.  The first is a `highly-clustered'
model, in which SCUBA sources cluster as strongly as the Extremely Red
Object (ERO) population. These galaxies have the strongest $w(\theta)$
of any known population at $z>1$ and as such represent a realistic
upper-bound to the strength of clustering expected from the SCUBA
sources (particularly since the SCUBA sources are believed to have a
much broader $n(z)$ distribution, which should dilute the angular
signal). We simulate a population with a standard 2-point angular
correlation function: $w(\theta) = (\theta/\theta_0)^{-0.8}$, where
the clustering strength for EROs is given by $\theta_0=1.5\times
10^{-2}$ deg (Daddi et al. 2000; Roche et al. 2002).  The
`least-clustered' model has an amplitude $\theta_0=0$, equivalent to a
random Poisson distribution. We also simulate a population with an
`intermediate' level of clustering, with $\theta_0=7.5\times 10^{-3}$
deg.

Large mock catalogues covering several thousand sq degrees were
generated.  Tests confirmed that the mock catalogue generation
successfully reproduces a distribution with the desired $w(\theta$).
Independent regions were then selected and overlaid over the galaxy
distributions in the real fields.  The density histogram analysis was
then repeated for each set of mock data.  The fraction of sources
lying in regions of above-average galaxy density ($\rho > \bar{\rho}$)
is then calculated.  Normalizing these results to the actual number of
SCUBA sources in each field, we can then predict the likelihood
function for $N$ sources (from a total of $39$) to lie in regions of
above-average density.  The results for the 3 fields combined are
shown in Figure 6, based on 10,000 simulations for each value of
$\theta_0$. We note that the expectation value ($\simeq 16$) differs
from $39/2=19.5$, reflecting the fact that the galaxy population
itself is clustered. The typical density sampled by a small aperture
should be slightly below the average field density.

As expected, the simulations with clustered submm sources give a
broader likelihood distribution for $N$. A simulation with strong
clustering is more likely to place a large number of submm sources
together in an overdense region, or in a void. Nevertheless, the
probability of observing $N\geq 24$ is still very small, regardless of
the clustering strength. For a Poisson distribution $P(N\geq
24)=0.08$\%. For the intermediate clustering model $P(N\geq
24)=0.19$\%, rising to $P(N\geq 24)=0.31$\% in the case of strong
clustering. In conclusion, the probability of these results being
produced by chance correlations is $<0.4$ per cent {\em even if submm
sources cluster as strongly as EROs}.

We can also use the distribution in Figure 6 to estimate a lower limit
on the number of submm sources associated with low-z structure. We
estimate this by removing sources in overdense regions until the
probability of a chance alignment rises above $5$\%. In the case of
Poisson clustering, this occurs for $N<20$ ($4$ sources removed) while
in the strong clustering model this occurs at $N<21$ ($3$ sources
removed). Thus a conservative lower limit to the number of ``excess''
SCUBA sources associated with low-z structure is $3/39$ ($8$ per
cent) at the $95$ per cent confidence level.

\begin{figure}
\centering
\centerline{\epsfxsize=8truecm 
\figinsert{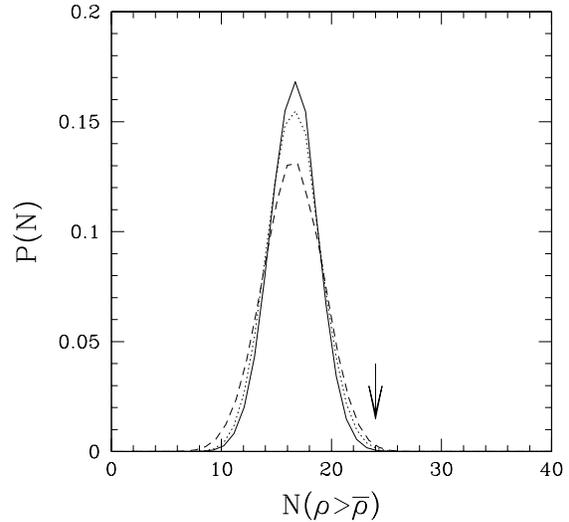}{0.0pt}}
\caption{Likelihood functions for the number of submm sources (from a
total of 39) expected to lie in overdense galaxy regions. Three models
are used to simulate the distribution of submm sources.  The solid
line assumes a Poisson distribution, the dotted line represents a
model with moderate angular clustering while the dashed-line is based
on simulations of strongly clustered sources (see Section 4). The
arrow shows the observed number ($N=24$) found for the
combined data in the 3 fields.}
\end{figure}

\section{Gravitational lensing bias}

We have presented evidence for a significant angular correlation
between SCUBA submm sources and overdensities of optical galaxies at
low-redshift. In this section we argue that gravitational lensing bias
provides a natural explanation for these effects, motivated by the
particularly steep submm source counts.

In a small solid angle, where the lensing magnification is given by
$\mu$, one can readily demonstrate that a population with cumulative
number counts given by a power law of index $\beta$ will be modified
as follows:

\begin{equation}
N'(>S) = \mu ^{\beta -1} N(>S)
\end{equation}

Observed submm number counts have a slope with $\beta \simeq 2.5$,
possibly steepening at $S_{850\mu{\rm m}}>8$mJy (Scott
et. al. 2002). Foreground large scale structure (e.g. galaxy groups
and overdensities) can readily lead to a magnification of $\mu \simeq
1.1-1.3$ (Hamana, Martel \& Futamase 2000).  Thus in the vicinity of
such structure one would {\em expect} an enhancement in number counts
of $15-50$ per cent.  Calculating the full effects of lensing bias
would require integrating over the full range of flux magnifications,
allowing for the relative positions of sources and lenses, so we refer
the reader to existing detailed models. These models predict that
between a few per cent (Blain et al. 1999) and 30 per cent (Perotta et
al. 2003) of submm sources at $S_{850\mu{\rm m}}>10$mJy will only be
seen because of lensing bias, with the precise fraction depending
strongly on the bright-end slope of the submm luminosity
function. These predictions are consistent with our findings.

\begin{figure}
\psfig{figure=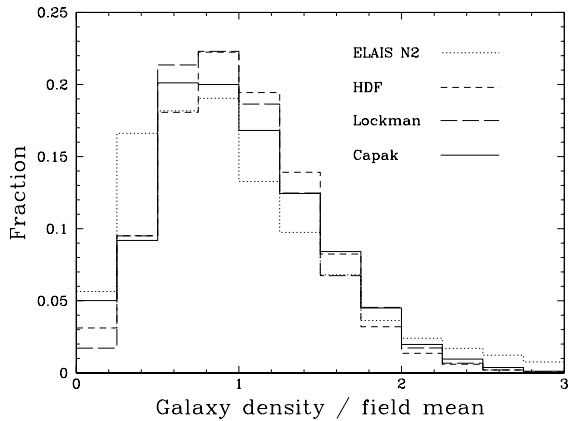,width=0.5\textwidth,angle=270}
\caption{Histogram of galaxy densities in the 3 submm fields to the
optical depths described in Section 2.2, compared with those from a
larger 0.2 square degree region from Capak et al. (2004).  Histograms
are based on 100,000 randomly placed apertures of 30 arcsec radius.}
\end{figure}

\section{Is ELAIS N2 an unusual field?}

The ELAIS N2 field shows the most dramatic correlation between submm
sources and low-redshift galaxies. In this section we investigate the
possibility that this may be an unusual field.

In Figure 7 we compare the density histograms in the three fields,
along with the distribution obtained from a much larger 0.2 sq degree
region from the Capak et al. (2003) HDF Subaru catalogue (see Section
2.2).  This shows that the ELAIS N2 field does have a broader
distribution of galaxy densities, consistent with a more strongly
clustered population.  An investigation of the galaxy-galaxy
auto-correlation function confirms that this is the most strongly
clustered field, with an auto-correlation amplitude approximately $40$
per cent higher than an average blank field at $R<23$ on 1-arcmin
scales (Roche et al. 1996).  From the compilation of clustering
measurements given in Roche et al. (1996) we estimate that
approximately $5-10$ per cent of blank fields of this size would show
galaxy clustering comparable to (or stronger than) ELAIS-N2 at
$R<23$. The galaxy number counts appear broadly normal, although they
are approximately $2\sigma$ higher than average around $R=20-21$.

In summary, the ELAIS-N2 field does not appear to be particularly
unusual, but it does contain denser low-z structure than an average field
(and the Lockman/HDF regions). We note that in the lensing model one
would expect a stronger biasing signal in high density regions, and
our results are certainly consistent with such an interpretation.  In
turn, this could also explain the puzzling cross-correlation of submm
and X-ray sources in this field (Almaini et al. 2003).  The hard X-ray
sources are strong tracers of peaks in the density field at $z<1$
(Gilli et al. 2003; Elbaz \& Cesarsky 2003), and could therefore be
tracing the foreground lensing structure.

On balance, however, we urge caution in over-interpreting these data,
particularly given the relatively small number of submm sources per
field. An investigation using independent data over a much larger area
is clearly required to overcome sample variance and investigate
further. This may soon be possible with the SHADES survey
(http://www.roe.ac.uk/ifa/shades/) or the future generation of surveys
with SCUBA2 (Holland et al. 2003).

\section{Conclusions}

From an analysis of three bright, flux-limited surveys we find
evidence for a significant angular correlation between SCUBA submm
sources and low-redshift galaxies. In particular, the distribution of
submm sources appears to be skewed towards overdense regions in the
$R<23$ galaxy population, consistent with $20-25$ per cent of the
submm population being apparently associated with low-redshift
structure, although there are strong field-to-field variations.  In
contrast, recent determinations of the $n(z)$ distribution for bright
submm sources suggest that at most only $5$ per cent are found at
$z<1$ (Chapman et al. 2003).  We note that the submm sources used in
our analysis span a similar range in flux density to the work of
Chapman et al. (with strongly overlapping samples) so survey depth is
unlikely to explain the difference.

Simulations suggest that the significance of these findings does not
depend strongly on the intrinsic clustering strength of the submm
populations.  Even for a population with an angular correlation
function as strong as Extremely Red Objects (EROs), the probability of
obtaining these correlations by chance is less than $0.4$ per cent.
The simulations also provide a conservative lower bound of $8$ per
cent on the ``excess'' fraction of submm sources associated with
low-z structure (at the $95$ per cent significance level).

Interestingly, the signal appears to be dominated by the brightest
sources with a flux density $S_{850\mu{\rm m}}>10$ mJy, although a
larger sample will be required to investigate the precise relationship
as a function of submm flux.

We argue that these findings are consistent with the expectations of
gravitational lensing bias, which is particularly strong in the bright
submm regime due to the unusually steep source number counts.  These
effects will have consequences for attempts to measure the intrinsic
clustering of submm sources, since the correlation function
$w(\theta)$ will be enhanced by excess pairs in the vicinity of
low-redshift structure (Moessner \& Jain 1998). Large submm surveys
now underway, such at the SHADES survey may be able to disentangle
these effects by simply excluding sources which lie in the vicinity of
low-z structure. Such techniques might also be used to estimate the
true (unlensed) source number counts, which may be much steeper than
presently accepted if the lensing bias is substantial. This could
reflect a sharp turn over in the underlying luminosity submm
function, and hence in the mass function of galaxies at high-redshift
(Benson et al. 2003).

\section*{ACKNOWLEDGMENTS}

OA and DMA acknowledge the support of the Royal Society. C. Liu
acknowledges the support of the Department of Astrophysics and Hayden
Planetarium at the American Museum of Natural History.  We thank Chris
Wolf for providing details of the $n(z)$ distribution from the
COMBO-17 survey and Peter Capak for providing details of the HDF
Subaru imaging. We also grateful to Meghan Gray, Takashi Hamana, John
Peacock, Rob Smith and Colin Borys for helpful discussions and advice.

\appendix

\section{Monte-Carlo Simulations of 2-D clustered populations}

There are many ways to simulate a clustered galaxy population.  We
conduct our simulations using a modification of the halo model of
Peacock \& Smith (2000), adapting it to the 2-dimensional
regime. Specifically we wish to simulate a distribution of sources
with an angular correlation function of the form:

\begin{equation}
w(\theta) = (\theta/\theta_0)^{-\gamma}
\end{equation}

We assume that galaxies live in circularly symmetric halos of angular
radius $\phi_H$ with radial density profiles of the form $\rho \propto
r^{-\epsilon}$. By placing down halo centres at random it can be shown
(see Peacock \& Smith 2000) that the resulting population will have an
angular correlation function with slope $\gamma=2\epsilon-2$.  To
reproduce the standard observed slope of angular clustering
($\gamma=0.8$) we therefore require a halo density profile of the form

\begin{equation}
\rho \propto r^{-1.4}
\end{equation}

The halos are then populated with randomly placed galaxies according
to this density distribution. Correlated pairs only occur within a
given halo, so the amplitude of the correlation function will be
diluted by the number of randomly placed halos. It can be shown 
that a correlation function of the form given in Equation A1
can be produced if the number density of halos is given by:

\begin{equation}
N = 0.2266~ \phi_H^{-2}  (\theta_0/\phi_H)^{-\gamma}
\end{equation}

The number of galaxies per halo is then chosen to ensure the desired
surface density for the full Monte-Carlo simulation. The choice of
halo radius ($\phi_H$) is somewhat arbitrary, but to ensure that the
2-pt function has the correct form on the 30-arcsec scales probed by
our analysis we choose values of $\phi_H$ in the range $\phi_H\sim
3-10~\theta_0$, corresponding to 1-9 arcmin given the correlation
lengths used in Section 4.  Comparing these simulations we found that
the precise choice of $\phi_H$ made no difference to the output
density histograms.

Finally we ensure that the simulated box is sufficiently padded, with
halo centres placed up to an angular radius $\phi_H$ outside the
survey region to be simulated. Tests are then conducted to ensure that
the mock catalogues generated by this procedure have the desired
correlation function $w(\theta$).

\end{document}